\documentclass[11pt]{article}

\usepackage{amssymb}
\usepackage{amsmath}
\usepackage{amsthm}

    \newcommand{\Reals}{\it I\kern-.4emR}
    
    \newcommand{\Notin}{/\kern-.6em\hbox{$\in$}}
    \newcommand{\Notequiv}{/\kern-.6em\hbox{$\equiv$}}
    
    \newcommand{\Ceals}{\it I\kern-.65emC}
    \newcommand{\MM}{\it I\kern-.4emM}
    \newcommand{\NN}{\it I\kern-.4emN}
    \newcommand{\yy}{\it Y\kern-.8emY}
    \newcommand{\zz}{\makebox[.80em]{\it Z\kern-.46emZ}}
    \newcommand{\tzz}{\makebox[.80em]{\scriptsize\it Z\kern-.46emZ}}

\newcommand{\R}[0]{{\mathbb{R}}}

\def\C{{\mathbb{C}}}

\def\R{{\mathbb{R}}}

\def\mrd{{M_{\R^d}}} 
\def\lrd{{L_{\R^d}}}

\newcommand{\onemat}[0]{{\mathbf 1}}

\newcommand{\ket}[1]{|#1\rangle}
\newcommand{\bra}[1]{\langle #1|}

\newcommand{\scalar}[2]{\langle #1|#2\rangle}

\newcommand{\m}[0]{{\bf m}}

    \newtheorem{theorem}{Theorem}[section]
    \newtheorem{lemma}[theorem]{Lemma}
    
    \newtheorem{proposition}[theorem]{Proposition}
    \newtheorem{corollary}[theorem]{Corollary}
    
    \newtheorem{observation}[theorem]{Observation}
    \newtheorem{conjecture}[theorem]{Conjecture}
    \newtheorem{definition}[theorem]{Definition}
\vspace{.6cm}

\begin{document}

\title{Real Mutually Unbiased Bases}

\author{
P. Oscar Boykin$^1$,
Meera Sitharam$^2$,
Mohamad Tarifi$^2$,
Pawel Wocjan$^3$\footnote{corresponding author wocjan@cs.caltech.edu}\\
{\small 1. Dept. of Electrical and Computer Engineering,
	   University of Florida,
	   Gainesville, FL 32611}\\
{\small 2. Computer and Information Science and Engineering,
	   University of Florida,
	   Gainesville, FL 32611}\\
{\small 3. Institute for Quantum Information,
           California Institute of Technology,
	   Pasadena, CA}\\
}

\maketitle

\begin{abstract}
We tabulate bounds on the optimal number of mutually unbiased bases in
$\R^d$. For most dimensions $d$, it can be shown with relatively
simple methods that either there are no real orthonormal bases that
are mutually unbiased or the optimal number is at most either $2$ or
$3$. We discuss the limitations of these methods when applied to all
dimensions, shedding some light on the difficulty of obtaining tight
bounds for the remaining dimensions that have the form $d=16n^2$,
where $n$ can be any number. We additionally give a simpler,
alternative proof that there can be at most $d/2+1$ real mutually
unbiased bases in dimension $d$ instead of invoking the known results
on extremal Euclidean line sets by {\sc Cameron and Seidel}, {\sc
Delsarte}, and {\sc Calderbank et al}.
\end{abstract}



\section{Introduction}
Two orthonormal bases ${\cal B}$ and ${\cal B}'$ of the Hilbert space
$\C^d$ are called {\it mutually unbiased} if and only if the modulus
of the standard complex inner product is
\begin{equation}
|\scalar{\phi}{\psi}|=1/\sqrt{d}
\end{equation}
for all $\ket{\phi}\in {\cal B}$ and all $\ket{\psi}\in {\cal B}'$.
The problem of determining bounds on the maximum number $M_{\C^d}$ of
bases over $\C^d$ that are mutually unbiased is an important open
problem \cite{KW:05} which has received much attention
\cite{Ivanovic:81, WF:89,bbrv02,KR:03,GHW:04, Wocjan}. We refer the
reader to e.g.\ \cite{ACW:04} for an overview of known bounds.

The most common application of MUBs is found in quantum cryptography
where MUBs are the quantum states used in most QKD protocols
\cite{bruss98,BP:00,CBKG:02,br04i}.

The MUB problem has remained open for over 20 years. For example, it
is not even known whether:\\ $\bullet$ {\sl for all $M$ there is a
$d'$ such that $M_{\C^d} \ge M$ for all $d > d'$}?  In this
manuscript, we consider the problem of determining $\mrd$ the maximum
number of mutually unbiased bases over $\R^d$ ({\it real MUBs}),
defined analogous to the above.  In the real case, the problem appears
to have a somewhat different character: elementary arguments give
tight bounds on almost all $d$; and the answer to the question
$\bullet$ is easily seen to be negative.  Additionally, we are able to
directly construct real MUBs which cannot be extended to an optimal
set even using complex bases; which means in general, maximal sets of
MUBs are not necessarily optimal.

Obtaining real MUBs is a special case of the geometric problem of
obtaining ``extremally or uniformly distributed'' collections of
Euclidean line sets \cite{Calderbank} with arbitrary, prescribed
angles. Observe  that a
collection of MUBs 
is exactly a collection of orthogonal sets of
lines passing through the origin  such that any one of the
following equivalent statements holds: (i) the angle between any pair
of lines taken from different sets is the same; (ii) the 
minimum angle over any pair of lines taken from different sets is 
maximized; (iii) the maximum angle over any pair of lines taken
from different sets is minimized.
(This formulation of MUBs
applies to any Hilbert space when ``angle'' is replaced by 
``magnitude of inner product.'')

\begin{table}
\begin{center}
\begin{tabular}{l|l|c}
dimension $d$        & number of real MUBs $\mrd$ & Section \\ 
\hline\hline
$d$            & $\le d/2+1$ \quad shown in \cite{Delsarte,Calderbank}
		       and by our alternative proof &
                       \ref{calderbank} \\ \hline
$4 \not\vert\, d$    & $1$ & \ref{basic} \\ \hline
$4n$ with $n$ not square & $\le 2$ & \ref{basic} \\
                     & $=2$ iff HM of size $d$ exists & \\ \hline
$4 s^2$              & $\le 3$ & \ref{sec:square} \\
                     & $\ge 2$ iff HM of size $d$ exists \\ \hline 
$4^i s^2$            & $\ge$ {\it MOLS}$(2^i s)+2$ if  
                       HM of size $2^i s$ exists \quad shown in
		       \cite{Wocjan} & \ref{sec:square} \\ \hline
$4^i$                & $=d/2+1$ \quad shown in \cite{Cameron} & 
                       \ref{calderbank}
\end{tabular}
\caption{Here $s$ is any positive odd integer, and $i$ is any
non-negative integer; {\it MOLS}$(r)$ denotes the maximum number of
mutually orthogonal Latin squares of order $r$. Recall that a {\it
Hadamard matrix} (abbreviated by HM) is a square matrix of size $d$
with entries in $\{-1,1\}$ with pairwise orthogonal columns.}
\label{tab:results}
\end{center}
\end{table}

In this paper, we show upper and lower bounds on the number of real
MUBs in various dimensions using elementary methods or known results
\cite{Wocjan, Calderbank}. These results are summarized in Table
\ref{tab:results}. The paper is organized as follows: Section
\ref{basic} introduces a simple proof technique which we use
throughout this paper. We use a simple technique to find the number of
real MUBs in dimensions which are not divisible by four ($\mrd=1$) and
non-square dimensions ($\mrd \le 2$).  The remaining dimensions can be
written as $d=4^i s^2$ with $s$ odd and $i\ge 1$.  We consider such
square dimensions $4^i s^2$ in Section \ref{sec:square} where we find
an upper bound ($\mrd \le 3$) in the case of $i=1$, and we use a
previous result\cite{Wocjan} to show a lower bound in the case of
$i>1$.  Section \ref{calderbank} gives a simpler, alternate proof for
a previously known upper bound ($\mrd \le d/2 +1$).

Finally, in Section \ref{greedy} we consider the question of whether 
all
sets of real MUBs can be extended into an optimal set of (complex or
real) MUBs and find
that the answer is negative.

\section{Most $d$'s admit at most 1 or 2 real MUBs}
\label{basic}

In this section, we consider non-square dimensions and dimensions
not divisible by four.  The proofs use an elementary technique
which we will make use of in later sections.
We will repeatedly use the following standard canonical form 
for real MUBs.

\begin{observation}
\label{canonical}
If $B_1,B_2,\ldots,B_m$ are a collection of real MUBs, without loss of
generality, we can always choose $B_1$ to be the standard basis
(simply multiply all matrices by $B_1^{-1}$ from the left). Putting
the elements of $B_1$ as columns, we get the identity matrix
$\onemat$. By Equation (1), all the other $B_i$ are orthogonal and
unbiased to $B_1$ if and only if their basis elements are the columns
of a Hadamard matrix scaled by $\frac{1}{\sqrt{d}}$; and $B_i, B_j$
are pairwise unbiased if and only if $B_i^{-1}B_j = B_i^t B_j $ is a
Hadamard matrix scaled by $\frac{1}{\sqrt{d}}$.
\end{observation}

\begin{proposition}
\label{1mub}
If $4 \not\vert\, d$, then $\mrd = 1.$
\end{proposition}
\begin{proof}
Let $B_1$, $B_2$ be two real MUBs in $\R^d$.  Using the canonical form
of Observation \ref{canonical} for $B_1$ and $B_2$, it is a {\sl
simple folklore fact} that ``$d$ must therefore be $2$ or a multiple
of $4$ since otherwise Hadamard matrix of size $d$ cannot exist.''
(The famous Hadamard conjecture \cite{HSS:99} is that for every
multiple of $4$, there exists in fact a Hadamard).

We prove this simple folklore fact here, since many of the arguments
presented in this note are elementary variants of this proof.  This
proof is similar to the one found in \cite[page 74]{Stinson:04}.  We
show the stronger statement that no more than $3$ orthogonal vectors
in $\{-1,1\}^d$ can exist unless $d$ is a multiple of $4$.

Assume there are three vectors $v_1,v_2,v_3\in\{-1,1\}^d$ that are
orthogonal to each other. Observe that permuting the entries of the
three vectors according to any permutation does not change the inner
product of any pair of them. Also note that changing the sign of the
entries by multiplying the three vectors entry-wise with any column
vector does not change the inner product of any pair, nor does it
change the magnitude of any entry.

We use these two facts to assume without loss of generality, that the
first vector $v_1$ is the all ones vector $(+,+,\ldots,+)$.  The
second vector $v_2$ should be orthogonal to the first, so again
without loss of generality it can be chosen as the vector
$(+,+,\ldots,+,-,-,\ldots,-)$, with the first half $+$'s and the
second half $-$'s. This requires $d$ to be even and completes the
Hadamard conjecture for $d=2$. The third vector $v_3$ should be
orthogonal to the first and the second, hence without loss of
generality, it can be chosen to be of the form
$(+,\ldots,+,-,\ldots,-,+,\ldots,+,-,\ldots,-)$, with $4$ blocks of
$+$'s and $-$'s with alternating signs. This is due to the ability to
permute the entries in any way we like. Any permutation that fixes
$v_1$ and $v_2$ is allowed, so we permute $v_3$ such that there are
$m_{++}$ values of $+$ followed by $m_{+-}$ values of $-$ in the first
half of the vector.  In the second half of the vector there are
$m_{-+}$ values of $+$ followed by $m_{--}$ values of $-$. This gives
rise to $4$ independent equations:
\begin{eqnarray*}
m_{++} + m_{+-} &=& \frac{d}{2}\\
m_{-+} + m_{--} &=& \frac{d}{2}\\
\langle v_1, v_3\rangle = m_{++} - m_{+-} + m_{-+} - m_{--} &=& 0\\
\langle v_2, v_3\rangle = m_{++} - m_{+-} - m_{-+} + m_{--} &=& 0
\end{eqnarray*}
The above equations give rise to the unique solution $m_{\pm\pm} =
d/4$ and consequently $4\vert d$.
\end{proof}

This shows that there are two real orthonormal bases are that are
mutually unbiased whenever $4\not\vert\ d$. Next we consider
non-square dimensions of the form $d=4n$.

\begin{proposition}
\label{2mub}
If $4 \vert d$, but $d$ is not a square, then $\mrd \le 2.$ In this
case, $\mrd = 2$ if and only if a Hadamard of order $d$ exists.
\end{proposition}

\begin{proof}
Assume the contrary and let $B_1$, $B_2$ and $B_3$ three such
orthonormal bases that are mutually unbiased. We also use $B_1,B_2$
and $B_3$ to denote the corresponding real matrices in the standard
canonical form of Observation \ref{canonical}.

As in the proof of the simple folklore fact and Proposition \ref{1mub}
given above, we can always achieve without loss of generality that the
first column of $B_2$ is $\frac{1}{\sqrt{d}}(1,1,\ldots,1)$. This is
done as follows. If the $i$th entry of $B_2$ is $\frac{-1}{\sqrt{d}}$
then multiply the $i$th row of $B_2$ and the $i$th row of $B_3$ by
$-1$. Clearly, this does not change the absolute values of the inner
products between the columns vectors of $B_1$, $B_2$ and $B_3$, nor
does it change the orthogonality of any of them.

Now, let $\frac{1}{\sqrt{d}}(s_1,s_2,\ldots,s_d)$ (with $s_i=\pm 1$)
be the first column of $B_3$. Eq.~(1) applied to the first column of
$B_2$ and the first column of $B_3$ implies that
\begin{equation}
|\sum_{i=1}^d s_i| = \sqrt{d}\,.
\end{equation}
The absolute value of the sum is clearly a natural number, whereas
$\sqrt{d}$ is irrational if $d$ is not a square. Therefore, there
cannot exist a third orthonormal real basis $B_3$ that is mutually
unbiased to $B_1$ and $B_2$.

This shows that the maximum number of real MUBs is at most $2$ for
dimensions $d$ that are not squares.  However, if there is a Hadamard
matrix of order $d$, then in fact, $B_2$ exists, and we have exactly 2
real MUBs.
\end{proof}

In summary, we have handled the case of all dimensions not divisible
by $4$ by showing that there are not even two MUBs. In non-square
dimensions which are divisible by four, there are at most two MUBs.
In the next section, we will consider all other dimensions.

\section{Square dimensions divisible by $4$}
\label{sec:square}
We have so far shown that unless $d = 4m^2$ for some positive integer
$m$, there are at most 2 real MUBs, such dimensions can be written as
$d=4^i s^2$ with $s$ odd.  First we consider the case of $i=1$ and
then consider the case of $i>1$.

\begin{proposition}
\label{3mubs} 
If $d = 4s^2$ for an odd positive integer $s$, the $\mrd \le 3$;
furthermore $\mrd \ge 2$ provided that a Hadamard matrix of order $d$
exists.
\end{proposition}

\begin{proof}
We show a stronger Lemma \ref{2mulls} that involves mutually unbiased
{\it lattice lines} (as opposed to entire bases).  Using this lemma
and the canonical form of Observation \ref{canonical}, it follows that
only $2$ further MUBs exist that are unbiased to the standard
basis.
\end{proof}

\begin{definition}
A {\it lattice line} in $\R^d$ is a line passing through the origin
and some point $v$ in $\{-1,1\}^d$.  We denote the line by either of
the vectors $v$ or $-v$.  Two lattice lines $v$ and $w$ are {\it
mutually unbiased} if $|\langle v,w \rangle| = \frac{1}{\sqrt{d}}$,
where $\langle\cdot,\cdot\rangle$ is the standard inner product in
$\R^d$.  We denote by $L_{\R^d}$ the maximum number of mutually
unbiased lattice lines that can be found in $\R^d$.
\end{definition}

\begin{lemma}
\label{2mulls}
If $d = 4s^2$ for an odd positive integer $s$, 
then $\lrd \le 2$. 
\end{lemma}

\begin{proof}
Assume to the contrary that there are $3$ such lines represented by
the vectors $v_1,v_2,v_d\in\{-1,1\}^d$. Consider the vectors
corresponding to those lines.  As in the proof of the simple folklore
fact and Proposition \ref{1mub}, we assume that $v_1=(+,+,\ldots,+)$
is the all-one-vector and we consider a partition of these vectors
into $4$ blocks with lengths $m_{++},m_{+-},m_{-+}$, and $m_{--}$. We
obtain the following equations:
\begin{eqnarray*}
\langle v_i, v_i\rangle = m_{++} + m_{+-} + m_{-+} + m_{--} &=& 4s^2\\
\langle v_1, v_3\rangle = m_{++} - m_{+-} + m_{-+} - m_{--} &=& \pm 
2s\\
\langle v_1, v_2\rangle = m_{++} + m_{+-} - m_{-+} - m_{--} &=& \pm 
2s\\
\langle v_2, v_3\rangle = m_{++} - m_{+-} - m_{-+} + m_{--} &=& \pm 
2s\\
\end{eqnarray*}
One can solve for $m_{++}$ by adding all the equations together to
get:
\[
m_{++} = s^2 + \frac{s(\pm 1 \pm 1 \pm 1)}{2}
\]
Since the value in the parenthesis can only add up to an odd number
$(-3,-1,1,3)$, and due to the fact that $s$ is odd and the product of
two odds is odd, $\frac{s(\pm 1 \pm 1 \pm 1)}{2}$ cannot be an
integer. However, if $m_{++}$ is not an integer, we have a
contradiction, proving the result.
\end{proof}

Hence we see that if $d=4^i s^2$ with $s$ odd and $i=1$, there are at
most $3$ real MUBs.  Now we consider the other half of the remaining
cases, i.e., where $i> 1$.  In this case, an earlier construction of
MUBs using Latin squares immediately gives a lower bound.

\begin{proposition}
\label{latin}
\cite{Wocjan} If $d = 4^i s^2$, where $s$ is any positive integer,
then $\mrd \ge {\it MOLS}(2^i s) +2$, provided that there exists a
Hadamard matrix of order $2^i s$, where {\it MOLS}$(m)$ denotes the
maximum number of mutually orthogonal Latin squares of order $m$.
\end{proposition}

This construction is contained in Appendix \ref{ap:latinmub}.  Since
there is a Hadamard of order $2$, the above construction also works in
dimension $4$ and is optimal in that case. We obtain the following
three mutually unbiased bases:
\[
{\cal B}_1 := \left\{
\frac{1}{\sqrt{2}}\left(
\begin{array}{r}
1 \\
1 \\
0 \\
0 
\end{array}
\right)\,,
\frac{1}{\sqrt{2}}\left(
\begin{array}{r}
1 \\
-1 \\
0 \\
0 
\end{array}
\right)\,,
\frac{1}{\sqrt{2}}\left(
\begin{array}{r}
0 \\
0 \\
1 \\
1 
\end{array}
\right)\,,
\frac{1}{\sqrt{2}}\left(
\begin{array}{r}
0 \\
0 \\
1 \\
-1 
\end{array}
\right)
\right\}\,,
\]
\[
{\cal B}_2 := \left\{
\frac{1}{\sqrt{2}}\left(
\begin{array}{r}
1 \\
0 \\
1 \\
0 
\end{array}
\right)\,,
\frac{1}{\sqrt{2}}\left(
\begin{array}{r}
1 \\
0 \\
-1\\
0 
\end{array}
\right)\,,
\frac{1}{\sqrt{2}}\left(
\begin{array}{r}
0 \\
1 \\
0 \\
1 
\end{array}
\right)\,,
\frac{1}{\sqrt{2}}\left(
\begin{array}{r}
0 \\
1 \\
0 \\
-1 
\end{array}
\right)
\right\}\,,
\]
\[
{\cal B}_3 := \left\{
\frac{1}{\sqrt{2}}\left(
\begin{array}{r}
1 \\
0 \\
0 \\
1 
\end{array}
\right)\,,
\frac{1}{\sqrt{2}}\left(
\begin{array}{r}
1 \\
0 \\
0 \\
-1
\end{array}
\right)\,,
\frac{1}{\sqrt{2}}\left(
\begin{array}{r}
0 \\
1 \\
1 \\
0 
\end{array}
\right)\,,
\frac{1}{\sqrt{2}}\left(
\begin{array}{r}
0 \\
1 \\
-1\\
0  
\end{array}
\right)
\right\}\,.
\]

So far, we have seen that the upper bound on the number of real MUBs
in dimension $d$ is constant (either $1,2$ or $3$) unless $d=4^i s^2$
with $s$ odd and $i>1$. In the latter case, we found a lower bound on
the number of real MUBs following from the construction based on
mutually orthogonal Latin squares. In the next section we give an
upper bound on the number of real MUBs and see that the upper bound is
tight in dimension $d=4^i$.

\section{A general upper bound which is tight for $d=4^i$}
\label{calderbank}

In this section we consider general dimensions and give a new proof
that $\mrd \le d/2 + 1$.  Then we cite a construction for $d=4^i$
which obtains the upper bound, and thus we see that the bound is tight
in the case of $d=4^i$.

\begin{proposition}
\label{prop:delsarte}
\cite{Delsarte,Calderbank} We have the upper bound $\mrd \le d/2 +1$
for any $d$.
\end{proposition}

Proposition \ref{prop:delsarte} is a special case of the type of bound
proved by \cite{Delsarte} (using Jacobi polynomials) and
\cite{Calderbank} (using tensor algebra first principles), where they
consider the maximum number of lines that can be packed into $\R^d$
such that the angle between any pair of lines is one of two specified
angles.

Next, we give an alternative, simpler proof of this upper bound.  This
is based on an adaptation to $\R^d$ - of a result of \cite{bbrv02}
which equates the MUBs over $\C^d$ to so-called commuting classes of
unitary matrices of size $d\times d$.

\begin{theorem}
\label{thm:comclass}
\cite{bbrv02} There exists a set of $m$ MUBs in $\C^{d}$ if and only if
there are $m$ classes $C_1 ,\ldots, C_m$ of the following properties:
\begin{itemize}
\item each class $C_j$ consists of $d$ commuting matrices,
\item any two classes have only the identity matrix in common, and
\item all matrices in $C_1 ,\ldots, C_m$ are pairwise orthogonal with
respect to the trace inner product.
\end{itemize}
\end{theorem}
The corresponding MUBs $B_1,\ldots,B_m$ are the common eigenvectors of
the matrices of the commuting classes $C_1,\ldots,C_m$, respectively.

\begin{corollary}
\label{cor:realcomclass}
There exists a set of $m$ MUBs in $\R^d$ if and only if there are $m$
classes $C_1,\ldots,C_m$ with properties as in
Theorem~\ref{thm:comclass} with the additional property that all
matrices are real symmetric.
\end{corollary}

\begin{proof}
Since the eigenvectors of real symmetric matrices are real, it follows
from Theorem~\ref{thm:comclass} that there are $m$ real $MUBs$ under
the assumptions of the corollary. This proves one direction of the
equivalence.

For the other direction, we show how to construct $m$ commuting classes
$C_{1},\ldots,C_{m}$ of real symmetric matrices satisfying the
properties of Theorem~\ref{thm:comclass} starting from $m$ MUBs.

Let the $j$th MUB be given by
\[
B_j=\{\ket{\psi_1^j},\ldots,
\ket{\psi_d^h}\}\,.
\]
Assuming that $m\ge 2$, we can construct at least one real Hadamard
\[
H:=\sqrt{d} B_1^t B_2\,,
\]
where we use $B_1$ and $B_2$ to also denote the unitary matrices whose
column vectors are given by basis vectors of $B_1$ and $B_2$,
respectively. We may assume without loss of generality that the $B_1$
is the standard basis and that the first vector of $B_2$ is given by
the normalized. Then the first row of $H$ is the
all-one-vector. Denote the entries of $H$ by $h_{tk}$.

Then we define the $j$th commuting classes $C_j :=
\{U_{j,1},U_{j,2},\ldots,U_{j,d}\}$, where the matrices within the
class are given by
\[
U_{j,t}:=\sum_{k=1}^d h_{t,k} \ket{\psi_k^j}\bra{\psi_k^j}
\]
for $t=1,\ldots,d$ and $j=1,\ldots,m$.

Observe that the matrix $U_{j,t}$ is diagonal with respect to the
basis $B_j$ and that its eigenvalues are given by the entries of the
$t$th row of the real Hadamard matrix $H$. Therefore, theses matrices
are not only unitary but also real symmetric. Clearly the matrices
within each commuting class commute because they are diagonal with
respect to the same basis. The only matrix that is contained in all
commuting classes is $I=U_{j,1}$. It remains to show that they are all
orthogonal with respect to the trace inner product.

Observe that all matrices within a commuting class are orthogonal
because the rows of $H$ are orthogonal. This implies that all matrices
not equal to the identity matrix are traceless.

We can compute the inner product directly:
\begin{eqnarray*}
tr(U_{i,r}^\dagger U_{j,s}) & = &
tr(U_{i,r} U_{j,s}) \\
&=&
\sum_{x=1}^d\sum_{y=1}^{d}
h_{r,x} h_{s,y} 
|\langle \psi_x^i | \psi_x^j\rangle|^2\\
&=& 
\frac{1}{d} \sum_{x=1}^d \sum_{y=1}^d h_{r,x} h_{s,y}\\
&=& 
\frac{1}{d} tr(U_{i,r}) \, tr(U_{j,s}) = 0
\end{eqnarray*}
because one of the matrices is not the identity matrix. This
completes the proof.
\end{proof}

Now we are ready to give the alternative proof of Proposition
\ref{prop:delsarte}.

\begin{proof}
(of Proposition \ref{prop:delsarte}). If there are $m$ real MUBs, then
according to Corollary \ref{cor:realcomclass} we can construct $m$
classes of commuting real symmetric unitaries. Each class contains
$d-1$ traceless matrices.  Additionally, we can consider the identity
matrix, which is also a real symmetric unitary. All of these
$m(d-1)+1$ matrices are orthogonal.  The space of real symmetric
matrices is a vector space of dimension $d(d+1)/2$. The span of our
$m(d-1)+1$ matrices is a subspace of all real symmetric matrices,
thus:
\begin{eqnarray*}
m(d-1) + 1 &\le& \frac{d(d+1)}{2}\\
m &\le& \frac{d}{2}+1
\end{eqnarray*}

This constitutes a new proof for the upper bound of real $MUBs$ in
dimension $d$.
\end{proof}

A construction of $d/2+1$ line sets that correspond to real MUBs is
given in \cite{Cameron,Calderbank} for the special case of $d$ being a
power of 4.  The general upper bound of Proposition
\ref{prop:delsarte} shows that construction is optimal.

In the next section we consider the question of extending sets of MUBs
to try to reach the upper bounds of this section. Unfortunately, we
will see that such extensions in general are not optimal.

\section{Greedy methods do not work}
\label{greedy}

In this section, we consider the efficacy of greedy methods of
constructing real MUBs. We ask the question: {\sl Starting with an
arbitrary set of MUBs or mutually unbiased lattice lines, can one
extend the set to find an optimal number of MUBs or mutually unbiased
lattice lines?} Put another way: can we take any MUB that comes our
way, or must we be careful to choose sets that fit well together.

We consider the question: can the method of Proposition \ref{3mubs}
and Lemma \ref{2mulls} be extended to show an upper bound on the
number of real MUBs for $d = 4^is^2$ (to complement the lower bound of
Proposition \ref{latin})?

We could continue, without loss of generality, to partition $j$
vectors into $2^j$ blocks and write down equations that they must
satisfy in order to represent mutually unbiased lattice lines.
Showing the conditions on $j$ and $d$ under which such systems do not
have solutions would then give us an upper bound on $L_{\R^d}$ (and
$M_{\R^d}$) for specific dimensions $d$.  However, in this approach,
we end up with a family of linear Diophantine systems one for each $j$
and $d$, with inequalities and equalities, since the solutions
$m_{ij}$ (to the sizes of the blocks in the partition) should be
non-negative integers.  Showing the conditions on $j$ and $d$ for
which such a system does not have a solution does not appear to be
tractable.

Our approach is instead the following.  We construct a {\sl
particular} set of $2^i$ lattice lines in dimension $d=4^i s^2$ using
the $2^i\times 2^i$ Sylvester Hadamard \cite{HSS:99}.  We then show
that this particular set cannot be extended by even one lattice line.
If any set can be extended into an optimal set, then $\lrd = 2^i$ for
$d=4^i s^2$.  However, this result is not true, because we later give
a construction that gives $\lrd \ge 2^i(2^i s^2 - s)$ for some cases
of $i$, which shows that not every set can be extended into an optimal
set.

Let $H$ be the Sylvester Hadamard matrix of order $2^i$ \cite{HSS:99}
and denote its entries by $s_{kl}$ for $k,l=0,\ldots,N-1$, where
$N=2^i$. Let $w_0,w_1,\ldots,w_{N-1}$ be all-one-vectors of lengths
$b_0,b_1,\ldots,b_{N-1}$, respectively. The lengths will be
determined later. The vectors $v_0,v_1,\ldots,v_{N-1}$ have the
following block structure:
\begin{equation*}
v_k := 
(s_{k,0}\, w_0\,|\, s_{k,1} w_1 \,|\,\ldots\, |\, s_{k,N-1}\, w_{N-1})^t\,,
\end{equation*}
where $|$ means that we concatenate the sub-vectors $s_{ij}\, w_j$ and
$t$ denotes the transposition. The conditions for the vectors $v_k$ to
for a collections of lattice lines are given by the following equations
\begin{eqnarray*}
\langle v_k, v_l \rangle & = & 4^i s^2 \\
\langle v_k, v_l \rangle & = & 2^i s
\end{eqnarray*}
for $0\le i<j\le N-1$. It may appear that there are $2^i +
2^i(2^i-1)/2$ equations. But that is not so. Many of the equations are
equivalent to others. This is because the columns of the Sylvester
Hadamard matrix form a group\footnote{This group is isomorphic to
$Z_2\times \cdots\times Z_2$ (the $i$-fold direct product of the group
$Z_2$). The multiplication of the $k$th and $l$th column gives $m$th
column, where $m$ is obtained by XORING the binary numbers
corresponding to $k$ and $l$ and converting the resulting binary
number to decimal notation. We denote this operation by $m=k\oplus
l$.} under point-wise multiplication. So there are only $2^i$
equations, rather than the much larger number that appears above. More
precisely, if $\oplus$ denotes the group operation, then we have
\begin{equation*}
\langle v_k, v_l \rangle = \langle v_0, v_{k\oplus l}\rangle\,.
\end{equation*}
So, in fact we see the only equations we have to consider are of the
form 
\[
\langle v_0, v_0 \rangle = \sum_{j=0}^{N-1} b_j = 4^i s^2\,,\quad\quad
\langle v_0, v_k \rangle = \sum_{j=0}^{N-1} s_{k,j} b_j = \pm 2^i s\,,
\]
for $1\le k\le N-1$. In matrix notation, this means
\begin{equation*}
H 
\left(
\begin{array}{c}
b_0    \\
b_1    \\
\vdots \\
b_{N-1}\\
\end{array}
\right)
=
\left(
\begin{array}{c}
4^i s^2 \\
\pm 2^i s   \\
\vdots  \\
\pm 2^i s
\end{array}
\right)\,.
\end{equation*}
By multiplying both sides by $H^t/2^i$ we obtain the solution
\begin{equation*} 
\left(
\begin{array}{c}
b_0    \\
b_1    \\
\vdots \\
b_{N-1}\\
\end{array}
\right)
=H^t
\left(
\begin{array}{c}
2^i s^2 \\
\pm s   \\
\vdots  \\
\pm s
\end{array}
\right)
\end{equation*}
and it is clear that the resulting lengths $b_0,b_1,\ldots,b_{N-1}$
are integers. This completes the proof that there are at least $2^i$
lattice lines in dimension $d=4^i s^2$.

Next, we show that such a collection of Sylvester lattice lines cannot
be extended. Let $v_0,\ldots,v_{N-1}$ be a collection of $N=2^i$ such
lines in dimension $d=4^i s^2$. Assume there is vector $w$ that is
mutually unbiased to the vectors $v_0,\ldots,v_{N-1}$. We may assume
without loss of generality that in the $b_j$ block of $w$ consists of
a sub-block of $+1$'s of length $b'_j$ and a sub-block of $-1$'s,
where $b_j=b'_j+b''_j$. This follows from the fact that permuting the
entries of the vectors within the $j$th block according to the same
permutation does not change the vectors $v_0,\ldots,v_{N-1}$ and does
not change the inner product between $w$ and $v_0,\ldots,v_{N-1}$.
This leads to the following equation system:
\begin{eqnarray*}
(I_2\otimes H_{2^i})
\left(
\begin{array}{c}
b'_0 + b''_0 \\
\vdots \\
b'_{N-1} + b''_{N-1} \\
b'_0 - b''_0 \\
\vdots \\
b'_{N-1} - b''_{N-1}
\end{array}
\right)
& = &
\left(
\begin{array}{l}
H_{2^i} \otimes (1,\,\,\,\, 1) \\ \\
H_{2^i} \otimes (1,-1)
\end{array}
\right)
\left(
\begin{array}{c}
b'_0     \\
b''_0    \\
\vdots   \\
b'_{N-1} \\
b''_{N-1}
\end{array}
\right) \\
& = &
H_{2^{i+1}}\, (b'_0,b''_0,\ldots,b'_{N-1},b''_{N-1})^t \\
& = &
P\, (4^i s^2, \pm 2^i s,\ldots,\pm 2^i s)^t\,,
\end{eqnarray*}
where $P$ is a permutation matrix. Multiplying by
$H_{2^{i+1}}^t/2^{i+1}$ on both sides we obtain
\begin{equation*}
b'_0 = \pm 2^{i-1} s^2 \pm \frac{s(\pm 1\pm 1\ldots\pm 1)}{2}=
2^{i-1} s^2 \pm \frac{s k}{2}
\end{equation*}
for some odd number $k$ corresponding to the sum in the
parenthesis. But since $s$ is also odd the resulting number $b'_0$ is
not an integer. This is a contradiction completing the proof that the
collection of $2^i$ Sylvester lines cannot be extended.

Unfortunately, it turns out that for $d=4^{i}s^{2}$, our greedy
construction gave $2^i$ unbiased lattice lines, but we can do a lot
better for $i>1$ as shown below.

Specifically, in the (generalized) construction of Lemma \ref{2mulls}
given above, while the partitioning and choices of the first three
vectors (needed for $i=1$) are without loss of generality, the choice
of the 4th and further vectors (needed for $i\ge 2$) is specific and
loses generality.  Therefore, non-extendibility does not imply
optimality, except for $i=1$.

\begin{proposition}
\label{greedymulls}
We have $\lrd \ge d - \sqrt{d}$ provided that there is a Hadamard
matrix of order $d - \sqrt{d}$.
\end{proposition}

\begin{proof}
Assume that there is a Hadamard matrix $H$ of order $d-\sqrt{d}$. We obtain
$d-\sqrt{d}$ lattice lines in dimension $d$ by appending to the row
vectors of $H$ the all-one-vector of length $\sqrt{d}$.

In other words, if the $h_k$ is the $k$th row vector of $H$, then the
$k$ lattice vector has the form
\[
\vec{v}_k =(h_{k,1},h_{k,2},\ldots,h_{k,(d-\sqrt{d})},1,1,\ldots,1)\,,
\]
It is obvious that these vectors are lattice lines.
\end{proof}
The above proposition along with many examples gives counter-examples
to the greedy conjecture.  For instance consider $d=4^2$.  If any
maximal set were optimal, then $\lrd \le 2^2$.  But since there is a
Hadamard of dimension $12$ ($d - \sqrt{d}$), we see that $\lrd \ge
12$.

So far, we have shown that greedy methods for constructing mutually
unbiased lattice lines do not work. Next, we consider the efficacy of
greedy methods of constructing general (complex) MUBs. Specifically,
we give a negative answer to the question: {\sl is every maximal, or
non-extendible set of MUBs necessarily optimal?}

We consider the general, complex version of the MUBs constructed in
Proposition \ref{latin} \cite{Wocjan} based on Latin squares, we call
these the {\it Latin} MUBs: note that this construction yields at most
$\sqrt{d}+1$ MUBs for dimensions of the form $d=p^{2e}$, where $p$ is
a prime and $e$ is an integer, whereas in prime power dimensions it is
known that there are in fact $d + 1$ MUBs \cite{bbrv02}. Nets and the
Latin MUB construction are reviewed in Appendix \ref{ap:latinmub}.  We
obtain restrictions on possible extensions to latin MUBs in general and 
show that the set of latin MUBs in $d=4$ is maximal, i.e., not extendible, 
thus answering the above question in the negative.

\begin{observation}
\label{ob:ks-space}
The $s(s+1)$ incidence vectors of a $(s+1,s)$-net span $\R^{s^2}$.
\end{observation}
\begin{proof}
For any $p\in\{1,\ldots,s^2\}$, there is exactly one incidence vector
in each parallel class that has $1$ as entry at position $p$. Sum all
these vectors. This gives the vector
$v_p=(1,\ldots,1,(s+1),1,\ldots,1)$, where the value $(s+1)$ appears
at position $p$. Now the $s^2$ vectors $v_p$ span $\R^{s^2}$ because
the $p$th standard basis vector $e_p$ is simply: $(1/((s+1)s^2) ((s^2
+ s - 1)v_p - \sum\limits_{q\ne p} v_q)$.
\end{proof}

\begin{lemma}
\label{lm:latin-genhad}
In dimension $d=s^2$, any new basis unbiased to a $(s+1)$ collection
of Latin MUBs corresponds to a Hadamard matrix. (Note the standard
basis is not unbiased to any Latin MUB).
\end{lemma}

\begin{proof}
Assume there is a $(s+1,s)$-net and let us consider the $(s+1)$ Latin
MUBs that are constructed based on this net. We show that any
orthonormal basis of $\C^d$ that is mutually unbiased to the Latin
MUBs defines a generalized Hadamard matrix of order $d$, that is, the
absolute values of all entries is $1/\sqrt{d}$.

Let us consider the $i$th incidence vector $m_{bi}$ of the block
$b$. This incidence vector gives rise to $s$ basis vectors of the
$b$th Latin MUB. Denote the support of $m_{bi}$ by $S_i$. By
construction the $s$ basis vectors contain a Hadamard matrix $H$ of
order $s$ as a sub-matrix if we look at the entries at positions
contained in $S_i$.

Let $y=(y_1,\ldots,y_d)$ be any vector of the new ONB and denote by
$\tilde{y}_i=(y_{i_1},\ldots,y_{i_s})$ the sub-vector of length $s$ of
$y$ defined by the support
$S_i=\{i_1,\ldots,i_s\}\subset\{1,\ldots,d\}$. Since $y$ is mutually
unbiased to the $s$ basis vectors obtained from $m_{bi}$, we have
\begin{equation}\label{eq:unbiased}
\frac{1}{\sqrt{s}} H \tilde{y}_i = \vec{u}
\end{equation}
where $u=(u_1,\ldots,u_s)$ is some vector in $\C^s$ with $|u_j|^2 =
\frac{1}{d}$ for $j=1,\ldots,s$. Since the matrix $\frac{1}{\sqrt{s}}
H $ is unitary, it follows that $\|\tilde{y}_i\|^2 = \|u\|^2$. We have
$\|u\|^2 = s/d = 1/s$ and consequently
\begin{equation*}
\sum_{j\in S_i} |y_j|^2 = \langle m_{bi},
(|y_1|^2,\ldots,|y_d|^2)^t\rangle = \frac{1}{s}\,.
\end{equation*}
The above equation is true for all incidence vectors $m_{bi}$ and
gives a system of linear equations for the squares of the absolute
values $|y_j|^2$.

Since the support of each incidence vector has cardinality $s$, one
solution to the above system is by choosing all entries of $y$ to have
the same magnitude, that is, $|y_j|^2 = \frac{1}{s^2}$. Due to
Observation \ref{ob:ks-space}, the system has full rank and thus this
solution is always a unique solution. This complete the proof that the
entries of all basis vectors of the new ONB have the same magnitude
and consequently that corresponds to a generalized Hadamard matrix.
\end{proof}


\begin{proposition}
The $3$ standard Latin MUBs in dimension $d=4$ are unextendible.
\end{proposition}

\begin{proof}
Assume there is a vector $y$ that is mutually unbiased to all vectors
of the three Latin MUBs. It follows from Lemma \ref{lm:latin-genhad}
that all its entries have the same modulus. Without loss of generality
we may assume that $y=(1,a,b,c)/2$ after factoring out a common phase
factor.

Writing the equalities $|1+a|^2=|1+b|^2=|a+b|^2=2$ that follow from
eq.~\ref{eq:unbiased} it is easily seen that there is no such $a$ and
$b$ satisfying these conditions. This shows that we cannot even find a
single vector that is unbiased to the Latin MUBs for $d=4$. This
completes the proof that the three Latin MUBs cannot be extended.
\end{proof}

In dimension $d=4$ if one allows complex MUBs, $5$ MUBs may be found. 
However, if we start with the three real Latin MUBs, we cannot even
find a fourth MUB, complex or real, unbiased to the previous three.
Hence it is clear that a set of MUBs which cannot be extended is not
necessarily optimal.

\section{Conclusion}
In this paper we have found upper and lower bounds on the number of
real MUBs in all dimensions.  In most dimensions, the problem is
solved.  Assuming the Hadamard conjecture is true, the only
interesting dimensions left are $d=4^i s^2$, where $s$ is odd.  In the
case of $d=4s^2$, assuming the Hadamard conjecture is true, the only
question is: does $\mrd = 3$ hold always, or only for certain values
of $s$.  Additionally, what is $\mrd$ for $d=4^i s^2$ with $i>1$ and
$s$ odd?  Assuming the Hadamard conjecture is true, this paper shows a
lower bound valid for all $i>1$.  It would be interesting to find an
example that exceeds that lower bound.

The work of M.S. and M.T. supported in part by NSF Grants EIA 02-18435
and CCF 04-04116. The work of P.W. supported in part by NSF Grant EIA
00-86038 and by the NSA under ARO contract number W911NF-05-1-0294 and
by the NSF under contract number PHY-0456720.

\appendix

\section{Nets and Latin MUBs}
\label{ap:latinmub}

We quote the definition of a net, which is used in the 
Latin MUB construction\cite{Wocjan}.
Note that nets are equivalent to orthogonal arrays and
traversal designs\cite{Stinson:04}.

\begin{definition}[Net]${}$\\
Let
$\{\m_{11},\ldots,\m_{1s},\m_{21},\ldots,\m_{2s},\ldots,\m_{k1},\ldots,\m_{ks}\}$
be a collection of $ks$ incidence vectors of size $d=s^2$ that are
partitioned into $k$ {\it blocks} where each block contains $s$
incidence vectors. The incidence vectors are denoted by $\m_{b i}$,
where $b=1,\ldots,k$ identifies the {\it block} and $i=1,\ldots,s$ the
vector within a {\it block}. If the incidence vectors satisfy the
following conditions we say that they form a $(k,s)$-net.
\begin{enumerate}
\item The supports of all vectors within one block are disjoint, i.e.,
\begin{equation}\label{eq:disjoint}
\m_{bi}^T\, \m_{bj} = 0 
\end{equation}
for all $1\le b \le k$ and all $1\le i\neq j\le s$.
\item The intersection of any incidence vectors from two different
blocks contains exactly one element, i.e.,
\begin{equation}\label{eq:one}
\m_{bi}^T\, \m_{cj}=1
\end{equation}
for all $1\le b\neq c\le s$ and all $1\le i,j\le s$.
\end{enumerate}
\end{definition}
Note that our definition of $(k,s)$-nets is in accordance with the
usual definition of nets in design theory \cite[page 172]{CD:96}.

Let $\m\in\{0,1\}^d$ be an incidence
vector of Hamming weight $s$ and ${\bf h}\in\C^s$ an arbitrary column
vector. Then we define {\it the embedding of ${\bf h}$ into $\C^d$
controlled by $\m$}, denoted by ${\bf h}\uparrow\m$, to be the
following vector in $\C^d$
\begin{equation}
{\bf h} \uparrow \m :=
\sum_{r=1}^s
{\bf h}[r] \, \ket{j_r}\,,
\end{equation}
where ${\bf h}[r]$ is the $r$th entry of the vector ${\bf h}$,
$\{j_1,j_2,\ldots,j_s\}$ the support of $\m$ with the ordering
$j_1<j_2<\ldots<j_s$ and $\ket{j_r}$ the $j_r$th standard basis vector
of $\C^d$. A less formal way to define this vector is: the first
non-zero entry of $\m$ is replaced by the first entry of ${\bf h}$,
the second non-zero entry of $\m$ by the second entry of ${\bf h}$,
etc.

This operation is best illustrated by a simple example:
\[
\m:=
\left(
\begin{array}{c}
1 \\
0 \\
1 \\
0 \\
0 \\
0 \\
0 \\
0 \\
1 \\
\end{array}
\right)\in\{0,1\}^9\,,
\quad\quad
{\bf h}:=
\left(
\begin{array}{c}
1 \\
\omega \\
\omega^2 
\end{array}
\right)\in\C^3\,,
\quad\quad
{\bf h} \uparrow \m:=
\left(
\begin{array}{c}
1 \\
0 \\
\omega \\
0 \\
0 \\
0 \\
0 \\
0 \\
\omega^2
\end{array}
\right)\in\C^{9}
\]

\begin{theorem}[Construction of MUBs]${}$\\
Let
$\{\m_{11},\ldots,\m_{1s},\m_{21},\ldots,\m_{2s},\ldots,\m_{k1},\ldots,\m_{ks}\}$
be a $(k,s)$-net and $H$ an arbitrary generalized Hadamard matrix of
size $s$. Then the $k$ sets for $b=1,\ldots,k$
\begin{equation}\label{eq:ourMUBs}
{\cal B}_b := \left\{
\frac{1}{\sqrt{s}} 
\big( 
{\bf h}_l \uparrow \m_{bi}
\big)\,\, |\,\, l=1,\ldots,s\,,\,\, i=1,\ldots, s
\right\}
\end{equation}
are $k$ mutually orthogonal bases for the Hilbert space $\C^d$.
\end{theorem}

\end{document}